# High-resolution 3T to 7T MRI Synthesis with a Hybrid CNN-Transformer Model

Zach Eidex[1,2], Jing Wang[1], Mojtaba Safari[1], Eric Elder[1,5], Jacob Wynne[1], Tonghe Wang[3], Hui-Kuo Shu[1,5], Hui Mao[4,5] and Xiaofeng Yang[1,2,5*]

[1]Department of Radiation Oncology, Emory University, Atlanta, GA
[2]School of Mechanical Engineering, Georgia Institute of Technology, Atlanta, GA
[3]Department of Medical Physics, Memorial Sloan Kettering Cancer Center, New York, NY
[4]Department of Radiology and Imaging Sciences, Emory University, Atlanta, GA
[5]Winship Cancer Institute, Emory University, Atlanta, GA

Email: xiaofeng.yang@emory.edu

**Running title**: 7T ADC Map Synthesis

**Manuscript Type:** Original Research

# ABSTRACT

7 Tesla (7T) apparent diffusion coefficient (ADC) maps derived from diffusion-weighted imaging (DWI) demonstrate improved image quality and spatial resolution over 3 Tesla (3T) ADC maps. However, 7T magnetic resonance imaging (MRI) currently suffers from limited clinical unavailability, higher cost, and increased susceptibility to artifacts. To address these issues, we propose a hybrid CNN-transformer model to synthesize high-resolution 7T ADC maps from multi-modal 3T MRI. The Vision CNN-Transformer (VCT), composed of both Vision Transformer (ViT) blocks and convolutional layers, is proposed to produce high-resolution synthetic 7T ADC maps from 3T ADC maps and 3T T1-weighted (T1w) MRI. ViT blocks enabled global image context while convolutional layers efficiently captured fine detail. The VCT model was validated on the publicly available Human Connectome Project Young Adult dataset, comprising 3T T1w, 3T DWI, and 7T DWI brain scans. The Diffusion Imaging in Python library was used to compute ADC maps from the DWI scans. A total of 171 patient cases were randomly divided: 130 training cases, 20 validation cases, and 21 test cases. The synthetic ADC maps were evaluated by comparing their similarity to the ground truth volumes with the following metrics: peak signal-to-noise ratio (PSNR), structural similarity index measure (SSIM), mean squared error (MSE). The results are as follows: PSNR: $27.0 \pm 0.9$ dB, SSIM: $0.945 \pm 0.010$, and MSE: $2.0\text{E-}3 \pm 0.4\text{E-}3$. Our predicted images demonstrate better spatial resolution and contrast compared to 3T MRI and prediction results made by ResViT and pix2pix. These high-quality synthetic 7T MR images could be beneficial for disease diagnosis and intervention, especially when 7T MRI scanners are unavailable.

# 1. INTRODUCTION

7 Tesla (7T) magnetic resonance imaging (MRI) offers superior spatial resolution and contrast compared to conventional 3 Tesla (3T) or other lower field strength scanners.[1] These advantages make 7T MRI ideal for tasks that require the differentiation of fine structures and subtle contrasts such as in small tumor and cerebral microbleed (CMB) detection (Figure 1),[2] the diagnosis of psychiatric disorders,[3] and MR spectroscopy.[4] However, 7T MRI is an emerging technology and is not yet widely available. In addition, 7T MRI is more susceptible to field inhomogeneities and susceptibility artifacts, and can cause increased patient discomfort such as vertigo and dizziness.[5]

Diffusion-weighted imaging (DWI), compared to structural MRI, is more sensitive to tumor volumes, ischemic stroke, and cerebellar abscesses by measuring water diffusion, a marker for tissue cellularity and damage.[6] Through multiple DWI scans, the relative differences in diffusion at different field directions (B-vectors) and strengths (B-values) can be quantified to create a single apparent diffusion coefficient (ADC) map. However, given the specific data acquisition scheme of DWI (single-shot or echo planar imaging) and the requirement of multiple scans, ADC maps are very susceptible to image artifacts.[7] The clinical utility of ADC maps generated at 7T field strengths compared to weaker fields increases due to the higher resolution and contrast that is achievable, albeit at the cost of potential for unwanted image artifacts. Therefore, there is interest in creating synthetic 7T ADC maps from existing 3T MRI scanners to improve image quality, while mitigating potential imaging artifacts and overcoming hardware scarcity.

Advances in natural image translation have facilitated significant progress in the field of intramodal MRI translation. Pix2pix is a generative adversarial network (GAN) based on U-Net, which includes a generator and discriminator balanced by an adversarial loss function which encourages realistic outputs.[8,9] However, Pix2pix is limited by its reliance upon convolutional layers and cannot capture long-range dependencies. Transformer-based models capture global context, effectively solving this problem.[10] To mitigate the computational demands of transformer-based networks, convolutional layers are frequently employed in shallower layers to capture fine details while transformers are deployed in the deeper abstract layers. The Residual Vision Transformer (ResViT) model is a state-of-the-art method for intramodal and multimodal MRI synthesis which incorporates this strategy.[11]

In this study, we propose the Vision CNN-Transformer (VCT) model based on the ResViT generator, modified to produce highly accurate 7T ADC maps. Design choices are made to encourage lower memory requirements and improve training efficiency.

We make the following contributions:

(1) This is the first attempt to produce synthetic 7T ADC maps from 3T MRI sequences.
(2) We implement flash attention and automatic mixed precision to improve training efficiency. Efficiency gains permit a simpler training scheme which avoids pre-training the convolutional layers, dedicating more epochs to training transformer layers.
(3) ADC maps have spatial dimensions (173 x 204 x 173) relative to most deep learning applications, which are optimized for square images. We modify the model to be compatible with any shape divisible by 16, avoid resizing which can introduce artifacts, and save computational resources by using the original image size with paddling (176 x 208).

# 2. METHODS

The proposed architecture is illustrated in Figure 1. The model produces synthetic 7T ADC maps from both 3T T1-weighted (T1w) and 3T ADC map axial slices input as separate channels. A convolution-based encoder and decoder learn localized features and are connected by an information bottleneck which encourages learning more abstract features. The information bottleneck has 9 layers connected with residual skip connections and do not change the dimensions of the feature maps. Two of these layers are transformer-based to allow for global context. The transformer layers share weights to reduce computational complexity.

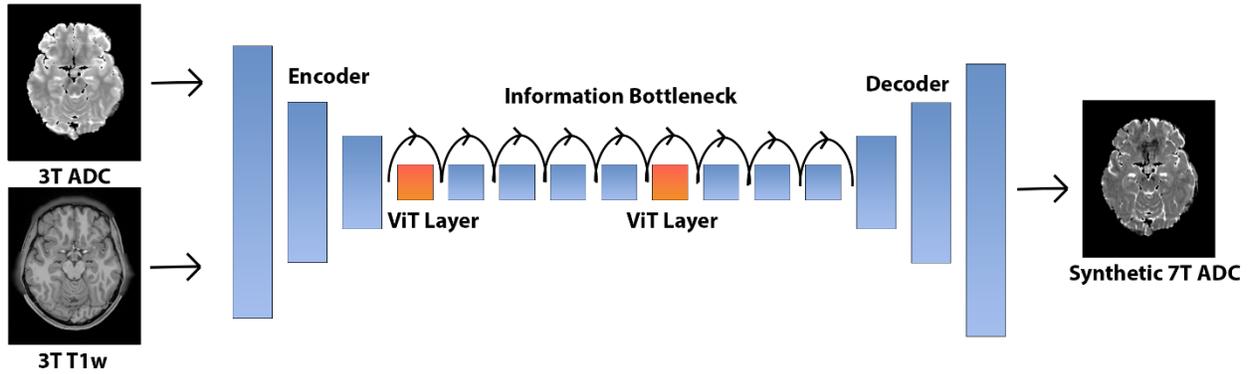

**Figure 1.** Schematic flow chart of the proposed algorithm. The orange boxes are the ViT blocks while blue boxes are convolutional layers. The Curved arrows represent residual skip connections which offer an alternate path bypassing the in-between layer to facilitate training a deeper network by backpropagating errors.

**2.1 Encoder and Decoder**

The encoder is composed of three convolutional blocks with the first having a large kernel size of 7x7 to capture broader image context. The following convolutional blocks are of kernel size 3x3 with a stride of 2. These layers further learn more abstract features and reduce the dimensionality by a factor of 4, easing the computational burden of the ViT blocks in the information bottleneck. The decoder is similarly comprised of three blocks. The first two are transposed convolutional blocks with a kernel size of 3x3 which restores the dimensionality back to the original input spatial dimensions. Finally, a 7x7 convolutional layer captures fine detail over a relatively large area.

**2.2 Information Bottleneck**

The information bottleneck is designed to encode abstract, long-distance features using both convolutional layers to reduce the computational cost and more powerful ViT blocks to boost performance and more accurately capture global context. Residual skip connections mitigate the vanishing gradient problem by allowing the gradient to bypass these layers during backpropagation.[12] The convolutional blocks shown in Figure 1 are comprised of two sequential 3 x 3 convolutional layers which are then connected to the input feature map through a skip connection. The ViT blocks are composed of two sequential 3 x 3 convolutional layers with a stride of two to reduce the dimensionality of the input feature maps by a factor of 4 prior to the transformer layer. After the transformer layer, two 3 x 3 transposed convolutional layers return the feature maps to their original spatial dimensions before passing through the ViT block. Finally, this output is connected to the input feature map through a residual skip connection.

ViT architectures have gained significant attention due to their potential to capturing long-range image context by across the entire feature map with a fully connected multilayer perceptron. However, ViT architectures introduce quadratic computational complexity that requires significant computational resources and training examples, rendering the primary challenge to be harnessing the transformer layers for efficient image synthesis.

Given limited MRI data, especially when compared to natural imaging tasks which are routinely trained on millions of images, and to encourage generalizability, pretrained weights from Google's R50+ViT-B_16 model trained on the ImageNet-21K were initialized in the ViT blocks.[10,13] The ImageNet-21K dataset contains approximately 14 million natural images and 21,000 classes allowing for highly generalizable features to be learned. The R50+ViT-B_16 model uses the ResNet-50 (R50) as a backbone with additional Vision Transformer layers to achieve state-of-the-art performance on the several datasets such as the cifar10, cifar100, and imagenet2012 datasets. Transfer learning was achieved by extracting the transformer component that expects 16x16 feature maps. These weights were then directly applied to the ViT blocks with the exception of the patch embeddings and positional encodings which were resized to 11x13 using bicubic interpolation to accommodate the input shapes of the feature maps after the convolutional layers (176 / 16 = 11, 208/ 16 = 13).

A significant reduction in computational complexity is achieved by placing the ViT blocks in the bottleneck region. In the encoder, the feature map size is reduced by a factor of 4. In addition, two 3x3 convolutional layers before the transformer layer in the ViT block further reduces the size by a total factor of 16. Further efficiency gains were realized by implementing a weight sharing strategy and by replacing the traditional attention mechanism with Flash Attention.[14] By assigning both ViT layers in the encoding and decoding branches with the same set of learnable parameters, the total number of trainable parameters in the model was significantly reduced by a factor of 41%, totaling approximately 123 million parameters. Finally, the attention mechanism of transformers is bottlenecked by read-and-write speeds to the GPU memory. By using tiling and recomputation to make use of the small, but extremely fast SRAM cache of the GPU, Flash Attention reduces the memory footprint and results in a significant speedup while being equivalent to the original attention mechanism.

## 3. Data Acquisition and Preprocessing

The Human Connectome Project (HCP) Young Adult dataset contains MRI scans of 1200 participants aged 22 to 35. Of these patients, 173 with both 3T and 7T DWI and 3T T1w MRI were included in this experiment. Both 3T and 7T DWI were converted to ADC maps using the Diffusion Imaging in Python (DIPY) library using the B-values and B-vectors provided in the dataset.[15] Because the generation of ADC maps is computationally intensive and the skull produces a strong, unwanted signal in ADC maps, a brain mask is needed. The masks provided with the dataset were atlas-based and found to be of poor quality. High quality brain masks were instead created *de novo* using SegSynth, an open source deep learning auto-contouring tool pre-trained on synthetic data.[16,17] To account for the difference of spatial resolution between 3T ADC maps (1.25 mm$^3$) and 7T ADC maps (1.05 mm$^3$), 3T ADCs were up-sampled to a resolution of 1.05 mm$^3$. The 3T T1w MRI had a resolution of 1.25 mm$^3$ and was reshaped using bicubic interpolation to match the shape of the 7T ADC maps. All images were rescaled to values in the range [0,1]. To combat very bright artifacts in the ADC maps caused by imperfect skull stripping, the upper-bound for the normalization of the ADC maps was set to the 99$^{th}$ percentile and values above it was clipped to the maximum value of 1. Finally, the normalized 3D ADC maps were split into 2D axial slices and saved in the portable network graphic (PNG) format.

## 4. Model Implementation and Performance Evaluation

### 4.1 Implementation Details

The VCT model was trained on a consumer-grade NVIDIA RTX 4090 GPU with 24 GB of memory. The dataset was augmented by randomly flipping the images in the coronal plane. An AdamW gradient optimizer (learning rate 2e-4, $\beta_1$ = .500, $\beta_2$ = .999, eps = 1e-6) was set to optimize the learnable parameters over 251 epochs or when the model no longer reduced the validation loss. For the hold-out test, training was stopped at 246 epochs.

The AdamW optimizer was chosen to minimize the loss function (L1 loss) for its improved generalization performance over the Adam optimizer due to a decoupling of the weight decay and gradient update.[18] 80 image slices were used for each batch.

### 4.2 Efficient Implementation

Improved efficiency, in terms of training speed and memory usage, makes training the computationally intensive ViT blocks in the VCT model over more epochs practical. The proposed method incorporates several technical strategies to boost efficiency, including automated mixed precision (AMP), Pytorch 2.0's torch.compile function, and training on inputs at their native resolution rather than upsampling to 256 x 256. AMP reduces memory consumption and accelerates training (assuming an optimized GPU architecture) by using half-precision values where possible while maintaining a sufficient population of full-precision values to support numerical stability. AMP thus allows for larger batch sizes and reduced memory usage without substantial loss in model accuracy.[19] PyTorch 2.0 introduced the torch.compile function as a means to further optimize PyTorch model execution. The function traces the model and converts it into a highly efficient intermediate representation optimized for computation. Finally, the ResViT model was modified to accommodate any input shape divisible by 16. ADC map inputs were originally 173 x 204 and were prepared for use by zero-padding to 176x208.

### 4.3 Validation and Evaluation

Model accuracy was evaluated with a hold-out test by randomly dividing the patients into training (120 patients: 16,434 slices), validation (21 patients: 2,876 slices), and testing (20 patients: 2,739 slices) segments. Five-fold cross-validation was also performed by randomly dividing the dataset into five subsets to assess model robustness and generalizability. The experiment were performed five times with each subset serving as the validation set once. In total, 136 patients comprised the training cohort and 35 patients comprised the validation cohort within each fold. Results were quantified using mean squared error (MSE), peak signal to noise ratio (PSNR), and structural similarity index (SSIM). Student's two-sided t-test was employed to compare model results with level of significance set at 0.05. MSE measures the voxel wise difference between the synthetic and ground truth volumes such that a value of zero means no difference.[20] PSNR is inversely related with the MSE so that higher PSNR values mean the synthetic images are more similar to the ground truth volume. Logarithmic scaling was applied to make it more closely align with human perception.[21] SSIM considers luminance, contrast, and structural similarity functions to most closely align with human perception. SSIM values range from -1 to 1 with 1 being perfect correspondence with the ground truth volume.[22] MSE, SSIM, and PSNR are defined below.

$$MSE = \frac{1}{n} \sum_{i=1}^{n}(X_i - Y_i)^2 \quad (1)$$

$$PSNR = 10 \log\left(\frac{MAX_I^2}{MSE}\right) \quad (2)$$

where $n$ is the total number of voxels, $X_i$ and $Y_i$ are the voxel intensity of the synthetic and ground truth volumes, and $MAX_I$ is the maximum possible voxel value of the ground truth volumes.

$$SSIM = l(x,y) \cdot c(x,y) \cdot s(x,y) = \frac{(2\mu_x\mu_y + 2C_1)(2\sigma_{xy} + C_2)}{(\mu_x^2 + \mu_y^2 + C_1)(\sigma_x^2 + \sigma_x^2 + C_2)} \quad (3)$$

where $l(x,y)$, $c(x,y)$, and $s(x,y)$ are the luminescence, contrast, and structural similarity functions respectively. $\mu_x$ and $\mu_y$ are the means of the synthetic and ground truth volumes, $\sigma_x^2$, $\sigma_y^2$, and $\sigma_{xy}$ are the variance of the synthetic volume, the variance of the ground truth volume, and the covariance between the synthetic and ground truth 7T ADC maps respectively. $C_1$ and $C_2$ are small constants to avoid dividing by zero (.01 and .03 respectively).

## 5. RESULTS

### 5.1 Model Performance

The synthetic 7T ADC maps generated by the VCT model are compared against ground truth 7T ADC maps as well as the synthetic ADC maps generated from the ResViT and pix2pix models. In addition, the similarity between 3T ADC maps and 7T ADC maps is measured to determine if the synthetic images show any improvement over 3T MRI. Example output images from 3 patients are shown in Figure 2 with zoom-in on regions. VCT most closely resembles the ground truth (GT) 7T ADC maps while pix2pix produces the worst outputs. The zoomed-in region of Figure 2c shows a case where a bright artifact was caused by the Synthseg auto-contouring model missing a small part of the skull. All methods successfully removed this artifact. Figure 3 provides a visual representation of the performance of the three models on the evaluation metrics (MSE, PSNR, and SSIM). Again, VCT performs the best, while pix2pix results in the least similar to ground truth 7T ADC maps with little overlap between the distributions. The similarity of the methods and the input 3T ADC maps to the ground truth volumes are additionally compared in Table 1. The VCT model achieves the best results based on the following evaluation metrics - MSE: .0015 ± .0004, PSNR: 28.1 ± 1.1, SSIM:.956 ± .009. These results show statistically significant improvement over competing methods, demonstrating p-values <.001 (two-tailed t test) for all metrics. Shown in Table 2, ablation studies were performed to determine the relative contributions of T1w MRI and 3T ADC maps on VCT's ability to generate accurate synthetic 7T ADC maps. The 3T ADC maps were more impactful to the performance of the model while using 3T T1w MRI only generated synthetic 7T ADC maps even less similar to 7T ADC maps than the 3T ADC maps. Finally, to verify the robustness of the VCT model, we preformed five-fold cross validation which showed almost identical results to the hold-out test - MSE: .0016 ± .0015 (95%), PSNR: 28.1 ± 1.2, SSIM: .954 ± .010.

### 5.2 Training Efficiency

To test the efficacy of the efficiency improvements, we perform ablation studies in Table 3. The most significant efficiency improvement was by using the original spatial dimensions with padding (176x208) compared to upsampling the images to 256x256 as was done to accommodate the required input shapes of pix2pix and ResViT. Flash attention and torch.compile showed marginal gains. AMP increased the training time per epoch by about 23 seconds which translates to almost 47 minutes over 120 epochs. However, AMP significantly increased the maximum batch size, so this was tolerated for increased flexibility. Compared to the original ResViT model, VCT is 45% faster while being able to use a much higher batch size.

## 6. DISCUSSION

In this study, we propose a VCT model for multimodal 7T ADC map synthesis which outperforms the ResViT and pix2pix models through improvements in training efficiency while retaining the multi-modal inputs and ViT layers which allows for maintenance of global image context and produces realistic outputs. These innovations enable improved performance realized by training over more epochs in the setting of limited computational resources. In addition, modifications are made to accommodate the unusual shape of the ADC map inputs (173 x 204 x 173). We further find that the inclusion of a discriminator in the pix2pix and ResViT models did not lead to improved performance over the VCT model. By generating accurate 7T ADC maps from 3T MRI, greater contrast and detail can be achieved without 7T MRI scanners, which currently are not widely available. These preferable image characteristics have shown 7T DWI to have an increased ability for tumor characterization and detection, so synthetic 7T DWI may also demonstrate superior clinical information.[23,24] 7T ADC maps were also found to mitigate artifacts common to both 7T MRI and ADC maps.

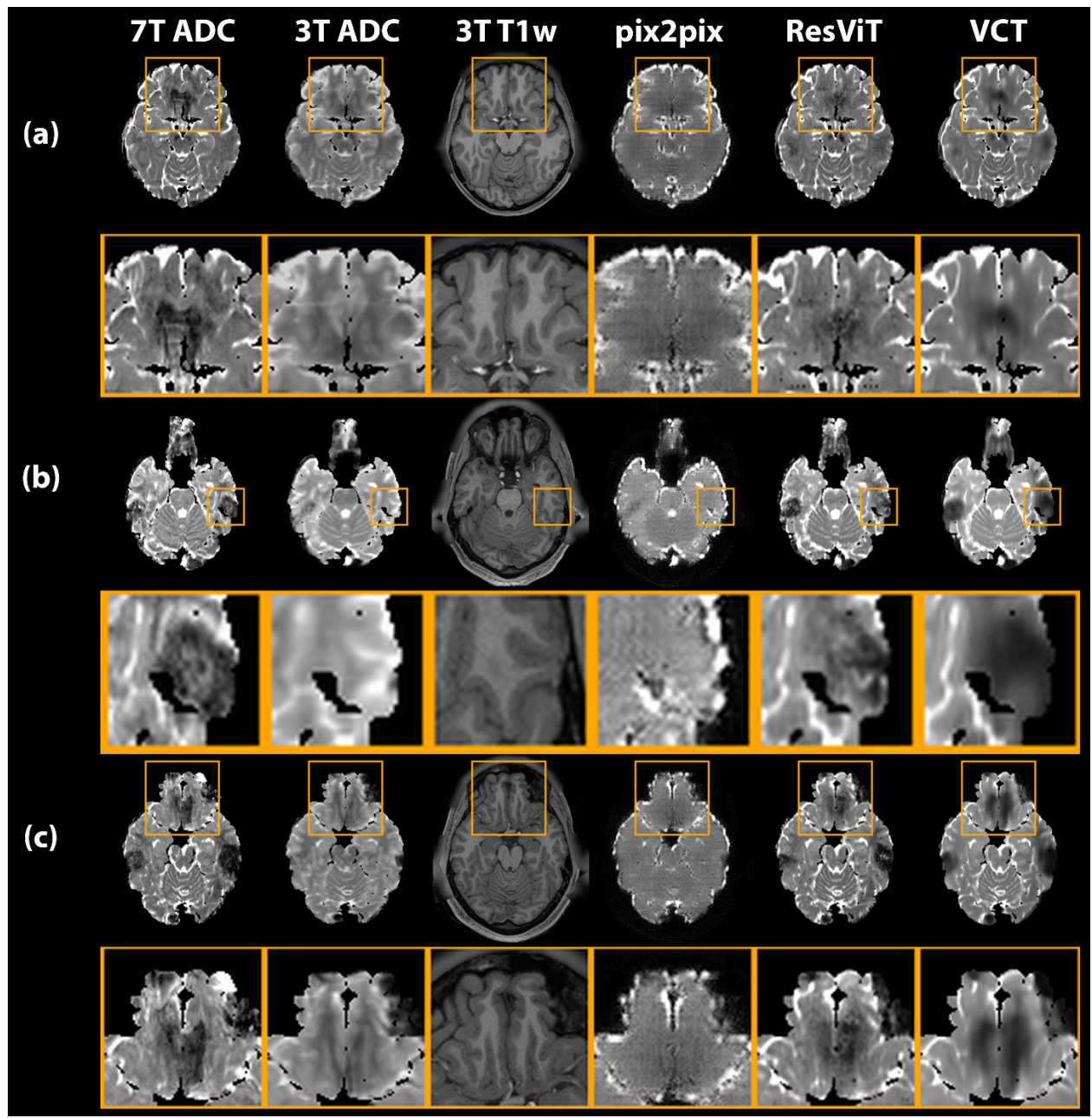

**Figure 2.** Example synthetic 7T ADC maps generated from pix2pix, ResViT, and VCT from 3 patients along with zoom-in on interesting regions. All methods took 3T ADC maps and 3T T1w MRI as inputs.

To our knowledge, this is the first work to synthesize 7T ADC maps from 3T sequences. Compared to ResViT and pix2pix, VCT achieves greater accuracy as measured by MSE, PSNR, and SSIM. Several studies have been published for mult-imodal, 7T, and super-resolution MR image translation tasks[25]. Modifying the StarGAN architecture for intramodal MRI synthesis, Dai et al generate four MR sequences (T1w, T2w, T1-contrast (T1c), and Fluid attenuated inversion recovery (FLAIR)) simultaneously from a single input sequence. Using T1w MRI as input, the SSIM for T1c, T2, and FLAIR were 0.974±0.059, 0.969±0.059, and 0.959±0.059 respectively. [26,27]. Xie *et al* used parallel CycleGANs to upscale T1w, T2w, T1c, and FLAIR MRI from a resolution of 1 x 1 x 3 to 1 x 1 x 1 voxels. They achieved an average SSIM value of 0.992 ± 0.004 for the T1w image translation compared to 0.927 ± 0.038. [28,29] This task is more straightforward than 3T ADC map to 7T ADC map translation, but shows a loss in image quality comparable to that observed here, due to bicubic image interpolation. Qu *et al* generated 7T T1w MRI from 3T MRI achieving a SSIM value of 0.8782 with a deep learning architecture, harnessing information in both the wavelet and spatial domains. Using a 15 patient dataset, they recovered a significant amount of detail and contrast found in the 7T T1w MRI.[30]

Comparing the input 3T T1w MRI and 3T ADC maps with the synthetic 7T ADC maps, the study reveals that the synthetic maps provide a significant advantage in terms of image quality and detail. Shown in Figure 2, the synthetic 7T ADC maps generated by the VCT model demonstrated substantial improvements over their 3T counterparts. We note that the normalized 3T ADC maps appear brighter than the normalized 7T ADC maps due to higher levels of noise and lower contrast in the 3T ADC maps. As verified quantitatively in Table 1, the VCT model outperforms ResViT and pix2pix, establishing a new state-of-the-art in all evaluation metrics (MSE, PSNR, and SSIM) and confirming that the synthetic 7T ADC maps are more closely aligned with the ground truth than the 3T ADC maps. Moreover, the VCT model demonstrated the ability to recover detail only visible in ground truth 7T ADC maps and remove skull artifacts in 7T scans.

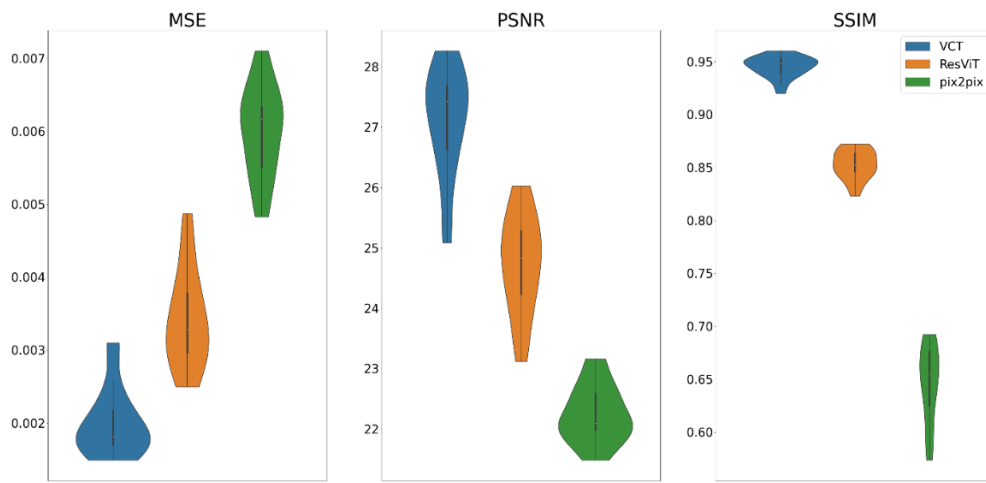

**Figure 3.** Violin plots of MSE, PSNR, and SSIM for the pix2pix, ResViT, and VCT models.

Regarding efficiency improvements, Table 3 demonstrates that the VCT model is significantly more efficient in terms of computational requirements compared to the ResViT model. Notably, the reduction in image size from 256 x 256 to 176 x 208 brought about impactful gains, allowing for lower memory requirements and faster training per epoch while avoiding bicubic interpolation in post-processing, which is known to compromise performance. In addition, AMP allowed for significantly larger batch sizes while only marginally increasing the training time per epoch. AMP typically reduces training time; however, this was likely observed in our case because the RTX 4090 GPU utilized here utilizes 16,384 compute unified device architecture (CUDA) cores optimized for single- and double- precision, but only 512 tensor cores optimized for half-precision. Finally, the lack of a discriminator in the VCT model has a significant impact on reducing model complexity and improving training efficiency.

**Table 1.** Evaluation metrics of 3T ADC maps and synthetic 7T ADC maps compared against GT 7T ADC maps. p-values are are <.001 for all evaluation metrics

|          | 3T ADC          | pix2pix         | ResViT          | VCT             |
|----------|-----------------|-----------------|-----------------|-----------------|
| **MSE**      | .0091 ± .0019   | .0060 ± .0003   | .0036 ± .0006   | .0015 ± .0004   |
| **PSNR (dB)**| 20.5 ± 0.9      | 22.2 ± 0.6      | 24.5 ± 0.7      | 28.1 ± 1.1      |
| **SSIM**     | .912 ± .015     | .646 ± .035     | .850 ± .014     | .956 ± .009     |
| **p-value**  | <.001           | <.001           | <.001           | x               |

**Table 2.** Ablation studies measuring the relative contributions of T1w MRI, 3T ADC maps, and the stability improvements on model performance.

|           | VCT (T1w only)   | VCT (3T ADC only) | VCT            |
|-----------|------------------|-------------------|----------------|
| **MSE**       | 0.0391 ± .0029   | .0019 ± .0004     | .0015 ± .0004  |
| **PSNR (dB)** | 14.1 ± 0.3       | 27.3 ± 1.0        | 28.1 ± 1.1     |
| **SSIM**      | .600 ± .030      | .948 ± .009       | .956 ± .009    |

**Table 3.** The effect of various efficiency techniques on training time and maximum batch size. Not using AMP resulted in the fastest training time but sacrificed batch size. These times were measured during the second epoch. Torch.compile was slowest in the first epoch since it required additional initial processing.

|                           | ResViT | VCT (No torch.compile) | VCT (No flash attn) | VCT (No AMP) | VCT (256x256) | VCT  |
|---------------------------|--------|------------------------|---------------------|--------------|---------------|------|
| **Time per Epoch (Minutes)** | 7.67   | 4.51                   | 4.28                | 3.86         | 7.8           | 4.25 |
| **Max Batch Size**        | 50     | 80                     | 78                  | 45           | 38            | 79   |

We acknowledge several limitations of the present work. The VCT model presented here is trained on 2D slices and so does not directly capture the full 3D context of the input data. While the convolutional layers employed in VCT provide an efficient way to capture the fine details, transformer-based alternatives such as Swin Transformer are also available. The dataset was comprised of only healthy young adults, so it remains to be seen how the results will generalize to patients with imaging-observable age-related anomalies such as tumors or implanted medical devices. However, given the successful application of related image translation tasks in these settings, we are optimistic about the generalizability of this model. [31-33] We intend in future work to incorporate full 3D context, replace the convolutional layers with Swin transformer layers[34,35], explore diffusion models in place of the GAN architecture[36], and perform experiments on more diverse datasets as they are made available.

## 7. CONCLUSION

This study presents a hybrid CNN-transformer model designed for multimodal 7T ADC map synthesis. It demonstrates marked improvements in accuracy and efficiency compared to current state-of-the art methods. The proposed method shows great promise in improving the diagnostic capabilities of existing 3T scanners by enabling the generation of high quality synthetic 7T ADC maps.

## ACKNOWLEDGMENTS

This research is supported in part by the National Cancer Institute of the National Institutes of Health under Award Numbers R01CA215718, R56EB033332, R01EB032680 and P30 CA008748. This work was supported in part by Oracle Cloud credits and related resources provided by Oracle for Research.

**Disclosures**

The author declares no conflicts of interest.